\documentclass[conference]{IEEEtran}
\IEEEoverridecommandlockouts
\usepackage{cite}
\usepackage{amsmath,amssymb,amsfonts}
\usepackage{algorithmic}
\usepackage{graphicx}
\usepackage{float}
\usepackage{textcomp}
\usepackage{xcolor}
\usepackage{subfigure}
\usepackage{url}
\usepackage{algorithm}
\usepackage{algorithmic}
\usepackage{makecell}
\usepackage{hyperref}
\usepackage{fancyhdr}
\def\BibTeX{{\rm B\kern-.05em{\sc i\kern-.025em b}\kern-.08em
    T\kern-.1667em\lower.7ex\hbox{E}\kern-.125emX}}

\begin{document}

\pagestyle{fancy}

\fancyhead[C]{This paper has been accepted by IEEE International Conference on Communications (ICC) 2023.}

\title{Virtual Reality in Metaverse over Wireless Networks with User-centered Deep Reinforcement Learning}

\author{\IEEEauthorblockN{Wenhan Yu}
\IEEEauthorblockA{Interdisciplinary Graduate Programme\\Nanyang Technological University\\
wenhan002@e.ntu.edu.sg }
\and
\IEEEauthorblockN{Terence Jie Chua}
\IEEEauthorblockA{Interdisciplinary Graduate Programme\\Nanyang Technological University\\
terencej001@e.ntu.edu.sg }
\and
\IEEEauthorblockN{Jun Zhao}
\IEEEauthorblockA{School of Computer Science \& Engineering\\ Nanyang Technological University\\
junzhao@ntu.edu.sg }
}

\maketitle
\begin{abstract}
The Metaverse and its promises are fast becoming reality as maturing technologies are empowering the different facets. One of the highlights of the Metaverse is that it offers the possibility for highly immersive and interactive socialization. Virtual reality (VR) technologies are the backbone for the virtual universe within the Metaverse as they enable a hyper-realistic and immersive experience, and especially so in the context of socialization. As the virtual world 3D scenes to be rendered are of high resolution and frame rate, these scenes will be offloaded to an edge server for computation. Besides, the metaverse is user-center by design, and human users are always the core. In this work, we introduce a multi-user VR computation offloading over wireless communication scenario. In addition, we devised a novel user-centered deep reinforcement learning approach to find a near-optimal solution. Extensive experiments demonstrate that our approach can lead to remarkable results under various requirements and constraints.

\end{abstract}


\section{Introduction}
\textbf{Background.} Maturing technologies in areas such as 6G wireless networks~\cite{6G} and high-performance extended reality (XR) technology~\cite{XR} has empowered the developments of the Metaverse~\cite{wang2022survey}. One of the key developments of the Metaverse is highly interactive and immersive socialization. Users can interact with one another via full-body avatars, improving the overall socialization experience.


\textbf{Motivation.} Virtual Reality is a key feature of an immersive Metaverse socialization experience. Compared to traditional two-dimensional images, generating $360^{\circ}$ panoramic images for the VR experience is computationally intensive. However, the rendering and computation of scenes of high resolution and frame rate are still not feasible on existing VR devices, due to the lack of local device computing power. A feasible solution to powering an immersive socialization experience on VR devices is through \textit{computation offloading}~\cite{MEC}. In addition, the metaverse is a \textit{user-centric} application by design, and we need to place the user experience at the core of the network design~\cite{metaverse}. Therefore, we have to consider a multi-user socialization scenario, in which each user has a different purpose of use and requirements. This propels us to seek a more user-centered and oriented solution. 


\textbf{Related work.} In recent years, VR services over wireless communication have been thoroughly studied in many previous works. Chen \textit{et al.} studied the quality of service of a VR service over wireless communication using an echo state network~\cite{chenVR}. 
However, none of the previous works considered the varying purpose of use and requirements between users. Although MEC-based VR services are thoroughly studied, few of them considered a sequential scenario over wireless communication. Machine-learning-based approaches have been widely adopted to tackle wireless communication challenges~\cite{Domenico, yunlong}, and DRL has been proven to achieve excellent performance. Meng \textit{et al.}~\cite{mengmetaverse} addressed the synchronization between physical objects and the digital models in Metvaerse with deep reinforcement learning (DRL). This is due to the ability of DRL agents to explore and exploit in self-defined environments\cite{RLC}. However, there are no existing works which has designed a \textit{user-centered} and \textit{user-oriented} DRL method.

\textbf{Approach.} This paper proposes a novel multi-user VR model in a downlink Non-Orthogonal Multiple Access (NOMA) system~\cite{NOMA}. We designed a novel DRL algorithm that considers the varying purpose of use and requirements of the users. We re-design the Proximal Policy Optimization (PPO) algorithm~\cite{PPO} with a reward decomposition structure.


\textbf{Contributions.} Our contributions are as follows:

\begin{itemize}
\item \emph{User-centered Computation Offloading VR Formulation:} We study the user-centered Metaverse computation offloading over the wireless network, designing a multi-user scenario where a Virtual Service Provided (VSP) assists users in generating reality-assisted virtual environments.
\item \emph{HRPPO:} We crafted a novel DRL algorithm Hybrid Reward PPO (HRPPO) to tackle the proposed channel allocation problem. The HRPPO is imbued with the hybrid reward architecture (HRA), enabling it to have a more user-centred perspective.
\item \emph{DRL Scenario Design:}  The design of the three core DRL elements: state, action, and reward are explained in detail. Extensive experiments demonstrate the effectiveness of our method.
\end{itemize}

The rest of the paper is organized as follows. Section~\ref{models} introduces our system model. Then Section~\ref{sec:RLenv} and \ref{sec:algorithms} proposes our deep reinforcement learning approach and settings. In Section~\ref{experiment}, extensive experiments are performed, and various methods are compared to show the prowess of our strategy. Section~\ref{conclude} concludes the paper.

\begin{figure}[t]
\centering
\includegraphics[width=1\linewidth]{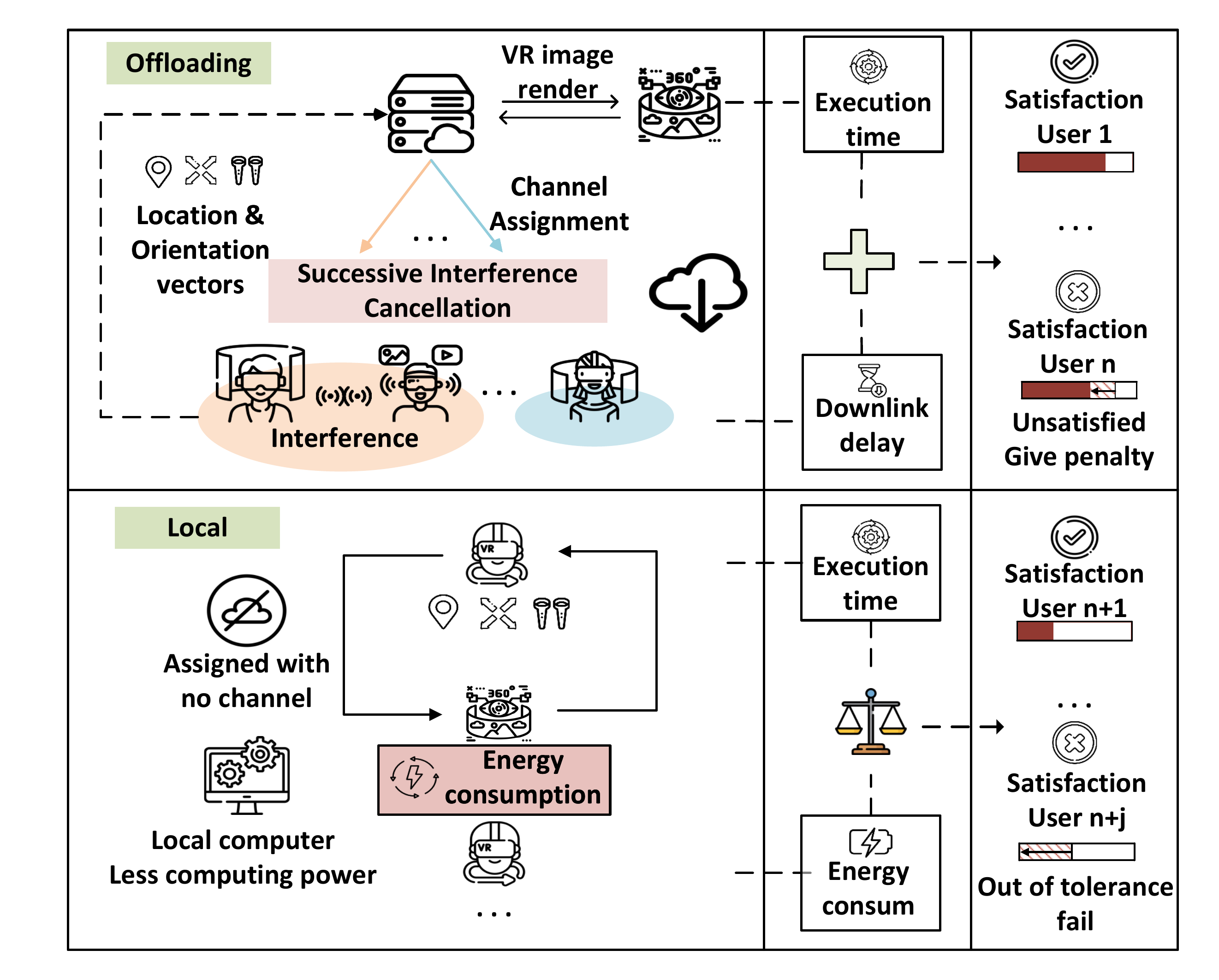}\vspace{-0.4cm}
\caption{Virtual reality in the Metaverse over wireless networks.}\vspace{-0.6cm}
\label{fig:model}
\end{figure}

\section{System model}
\vspace{-0.1cm}
\label{models}
Consider a multi-user wireless downlink transmission in an indoor environment, in which one second is divided into $T$ time steps. To ensure a smooth experience, we consider a slotted structure with the clock signal sent by the server for synchronization, and each slot contains one high resolution frame transmission, and the duration for each time slot is $\iota$ ($\iota=\frac{1}{T}$). In each time step, a sequence of varying resolution 3D scenes is generated by the VSP and sent to $n$ VR device users (VU) $\mathcal{N}=\{1,2,...,N\}$ with distinct characteristics (e.g., Computation capability) and different requirements (e.g., Tolerant delay). Each user is selected for either (1) \textbf{Computation offloading} by offloading the tracking vectors $\chi_n$~\cite{VRsurvey} to the virtual service provider (VSP) for scene rendering, or (2) \textbf{Local computing} by receiving scenes (tracking vectors) from others and render scenes locally with a lower computation capability and at the expense of energy consumption. If a user is selected for computation offloading, the VSP will generate the frame and send it back to them via a set of channels $\mathcal{M}=\{1,2,...,m\}$,

Each user can accept virtual scene frame rates as low as a minimum tolerable frames per second (FPS) $\tau_{n,F}$, which is the number of successfully received frames in a second. Considering that the tracking vectors are relatively very small~\cite{VRsurvey}, we assume that the vectors are transmitted with dedicated channels between VSP-VU and VU-VU, and neglect the overhead.

We use $\Gamma^t = \{\Gamma_1^t,\Gamma_2^t,...,\Gamma_N^t\}$ to denote the selection of downlink channel arrangement and inherently, the computing method (VSP or local computation). $\Gamma_n^t=m$ indicates VU $n$ is arranged to channel $m$ at time step $t$, and $\Gamma_n^t=0$ means user $n$ needs to generate locally. 

Thus, it is imperative to devise a comprehensive algorithm that takes into account the varying satisfaction threshold and requirements of the users. In the next section, we explain the computation offloading and local computing models in detail. The system model is shown in Fig.\ref{fig:model}.

\subsection{Computation offloading model}\label{model:off-Model}
We first introduce the computation offloading model based on the wireless cellular network. The server VSP will manage the downlink channels $\mathcal{M}$ of all VUs $\mathcal{N}$. And then, the server VSP Furthermore, we denote $D_{n}^{t}$ ($n \in \mathcal{N}$) as the size of the virtual scene frame at time step $t$ that needs to be transmitted to user $n$. 

We adopt the Non-Orthogonal Multiple Access (NOMA) system as this work's propagation model. In NOMA system, several users can be multiplexed on one channel by the successive interference cancellation (SIC) and superposition coding, and the received signals of VUs in channel $m$ are sorted in ascending order: $p_1|h_{1,m}^t|^2>p_2|h_{2,m}^t|^2>...>p_N|h_{N,m}^t|^2$~\cite{NOMA}. In this paper, we assume the decoders of the VUs can recover the signals from each channel through SIC. We denote $h_{i,m}^t$ as the channel gain between the VSP and the $n^{th}$ user allocated to channel $m$ at time step (iteration) $t$. The Downlink rate can be expressed as~\cite{NOMA}:
\begin{align}
    &{r}_{n}^{t}=W\log\left(1+\frac{{p}_{n}|h_{n,m}^{t}|^2}{\sum_{i=n+1}^{N}{p}_{i}|h_{n,m}^{t}|^2+W\sigma^2}\right).\label{eq:dlrate}
\end{align}

$P_d=\{p_1, p_2,...,p_N\}$ denotes the transmission power of each VU's device. Note that transmission power is not time-related in our scenario. $h_{n,m}^{t}=g_{n,m}^{t}l_n^{-\alpha}$ denotes the channel gain between VU $n$ and VSP in channel $m$, with $g_{n,m}^{t}$, $l_n$, $\alpha$ being the Rayleigh fading parameter, the distance between VU $n$ and VSP, and the path loss exponent, respectively. $W$ is the bandwidth of each channel, and $W \sigma^{2}$ denotes the background noise. Accordingly, the total delay $d_{n,o}$ of each frame in time step $t$ is divided into (1) Execution time and (2) downlink transmission time:
\begin{align}
    d_{n,o}^t = \frac{D_n^t\times C_n^t}{f_v} + \frac{D_n^t}{r_n^t}. \label{eq:offdelay}
\end{align}
where $f_v$ is the computation capability of VSP, and $C_n^t$ is the required number of cycles per bit of this frame~\cite{Csetting}. 

\subsection{Local computing model}\label{model:local-Model}
When VU is not allocated a channel, it needs to generate the virtual world frames locally at the expense of energy consumption. Let $f_n$ be the computation capability of VU $n$, and it varies across VUs. Adopting the model from~\cite{energy}, the energy per cycle can be expressed as $e_{n,cyc} = \eta f_n^2$. Therefore, the overhead of local computing in terms of execution delay and energy can be derived as:
\begin{align}
    &d_{n,l}^t = \frac{D_n^t\times C_n^t}{f_n}. \label{eq:ldelay}\\
    &e_{n,l}^t = \mu_n \times D_n^t \times C_n^t \times e_{n,cyc}. \label{eq:energy}
\end{align}
where $\mu_n$ is the weighting parameter of energy for VU $n$. The battery state of each VU can be different, then, we assume that $\mu_n$ is closer to $0$ with a higher battery.

\subsection{Problem formulation}
With the slotted structure, we set the VUs' maximum tolerable delay to be $\iota$ for every frame to be the problem constraint. Different users have different purpose of use (video games, group chat, etc.). Thus, they also have varying expectations of satisfactory number of frames per second $\tau_{n,F}$. We set the tolerable frame transmission failure count of VU $n$ as $\tau_{n,f}$. Initially, the tolerable frame transmission failure count of VU $n$ is defined as $\tau_{n,f}^0 = T-\tau_{n,F}$.
For each successive frame, the delay in exceedance of tolerable threshold leads to a decrease in VU's remaining tolerable count: $\tau_{n,f}^{t+1} = \tau_{n,f}^t-I_n^t$, where
\begin{align}
    I_n^t = 
    \begin{cases}
        1, &if~~d_{n,o}^t > \iota~or~d_{n,l}^t > \iota. \\
        0, &else.
    \end{cases}
\end{align}
Our goal is to find the near-optimal channel arrangement for the transmission of $T$ frames, to minimize the total frame transmission failure count and VU device energy consumption.
\begin{align}
\min\limits_{\Gamma^1,...,\Gamma^T}&\sum_{n \in \mathcal{N}}\sum_{t=0}^T
\left [ \omega_1 I_n^t+ \omega_2 e_{n,l}^t \right ].\\
s.t.~C1:&\tau_{n,f}^t \geq 0,~\forall n\in\mathcal{N}, \forall t\in[0,T]. \\
     C2:&\Gamma_n^t = \{0,1,...,M\},~\forall{n}\in\mathcal{N},\forall{t}\in[0,T].
\end{align}
The $\omega_1,\omega_2$ are the weighting parameters of delay and energy. Constraint $C1$ ensures that the frame transmission failure count of each user is within their tolerable limit. Constraint $C2$ is our integer optimization variable which denotes the computing method and channel assignment for each user at every time step.


This formulated problem is \textbf{sequential}, where the remaining tolerable frame transmission failure count $\tau_n^{f,t}$ of each user changes over time, and influences the following states. Thus, convex optimization methods are unsuitable for our proposed problem due to the huge space of integer variables and daunting computational complexity. Also, as the problem contains too many random variables, model-based RL approaches which require transition probabilities are infeasible techniques to tackle our proposed problem. We next introduce our deep RL environment settings according to the formulated problem.

\section{Deep reinforcement learning setting}
\label{sec:RLenv}
For a reinforcement learning environment (problem), the most important components are (1) State: the key factors for an agent to make a decision. (2) Action: the operation decided by an agent to interact with the environment. (3) Reward: the feedback for Agent to evaluate the action under this state. Thus, we expound on these three components next.

\subsection{State}
We included the following attributes into the state: (1) Each VU's virtual world frame size: $D_n^t$. (2) Each VU's remaining tolerable frame transmission failure count: $\tau_{n,f}^t$. (3) The channel gain of each VU: $h_{n,m}^t$. (4) The remaining number of frames to be transmitted at each time step: $(T-t)$.

 \begin{figure}[t]
    \centering
    \setlength{\abovecaptionskip}{-0.2cm}
    \includegraphics[width=0.75\linewidth]{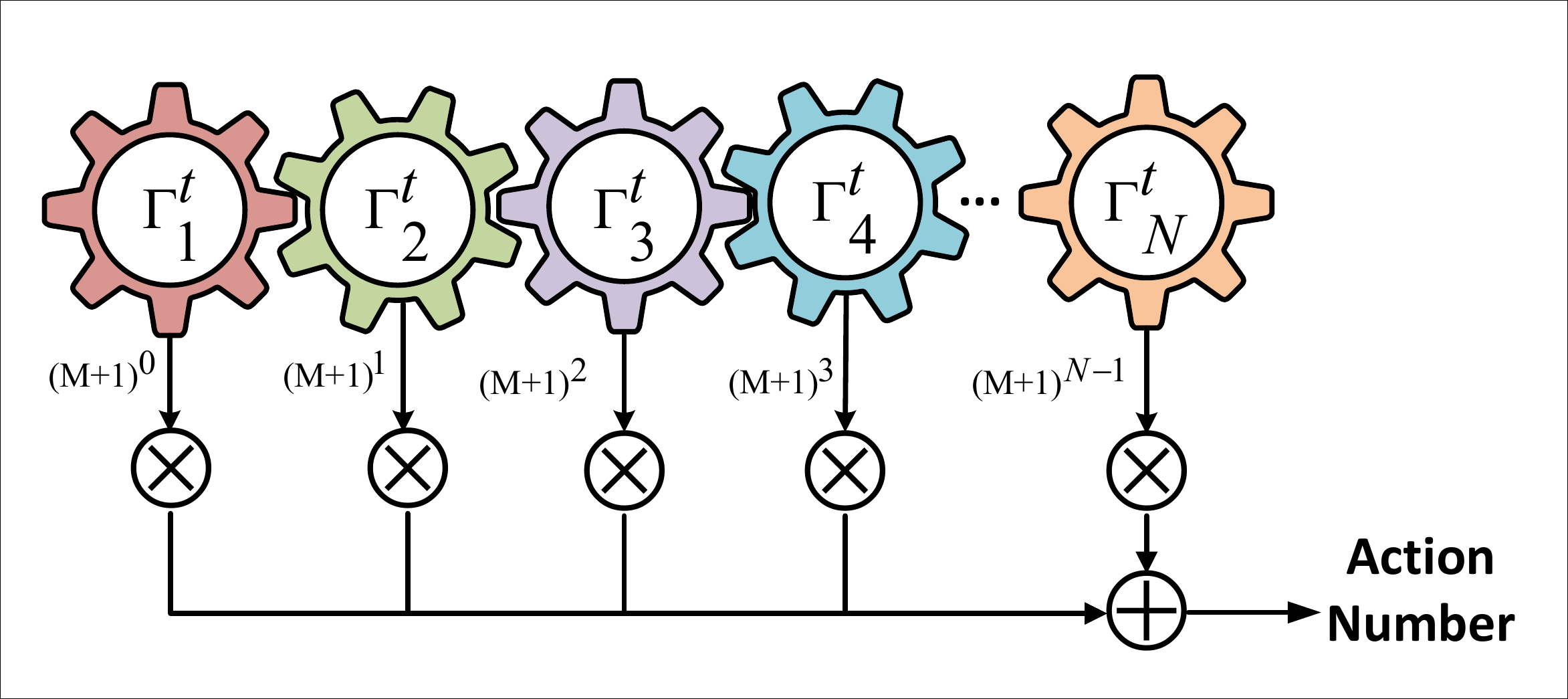}
    \caption{UL action encoding method.}
    \label{fig:actionencode}
    \vspace{-0.8cm}
\end{figure}

\subsection{Action}
The discrete action channel assignment to each VU is:
\begin{align}
    &a_u^t = \Gamma^t = \{\Gamma_1^t, \Gamma_2^t,..., \Gamma_N^t\}. \\
    &s.t.~~\Gamma^t_n\in \{0,1,...,M\}.
\end{align}
 In practice, we use a tuple in which there are $N$ elements corresponding to $N$ users and each element can take $M+1$ values, which corresponds to the number of channels; plus 1 for a user being assigned to perform local computing. However, we need to encode the discrete actions as discrete numbers to be evaluated by the neural network. The encoding method is shown in Fig.~\ref{fig:actionencode}.

\subsection{Reward}

As the main objective is to minimize the frame transmission failure counts and energy consumption, the overall reward $R_n^t$ for each VU contains: (1) a penalty $R_{n,f}^t$ for every frame transmission failure and (2) a weighted reward $R_{n,e}^t$ for energy consumption corresponding to VU's battery life. To implement the tolerance constraint $C1$, we give (3) a huge penalty $R_{n,end}^t$ corresponding to the number of frames left to be transmitted when any VU's remaining tolerable frame transmission failure count is $0$. In the circumstance of (3), the \textbf{episode ends immediately}.\\

\section{Deep Reinforcement learning Approach} \label{sec:algorithms}
Our proposed Hybrid Reward Proximal Policy Optimization (HRPPO) is based on Proximal Policy Optimization (PPO) algorithm, which is considered as the state-of-art RL algorithm~\cite{PPO}. HRPPO is inspired by the Hybrid Reward Architecture (HRA)~\cite{HRA}. Thus, PPO and HRA preliminaries will first be introduced. We will then explain the HRPPO.

\subsection{\textbf{Preliminary}}
\subsubsection{Proximal Policy Optimization (PPO)}
As we emphasize on developing a \textit{user-centred} model which considers VUs' varying purpose of use and requirements, policy stability is essential. Proximal Policy Optimization (PPO) by openAI~\cite{PPO} is an enhancement of the traditional policy gradient algorithm. PPO has better sample efficiency by using a separate policy for sampling, and is more stable by embedding policy constraint.

In summary, PPO has two main characteristics in its policy network (Actor): (1) \textit{Increased sample efficiency.} PPO uses a separate policy for sampling trajectories (during training) and evaluating (during evaluation) to increase sample efficiency as well. Here we use $\pi_\theta$ as the evaluating policy and $\pi_{\theta_{'}}$ as the data sampling policy. As we use the $\pi_{\theta_{'}}$ to sample data for training, the expectation can be rewritten as:
\begin{align}
    \mathbb{E}_{(s^t,a^t)\sim\pi_\theta}[\pi_\theta(a^t|s^t) A^t] &=  \mathbb{E}_{(s^t,a^t)\sim\pi_{\theta_{'}}}[\frac{\pi_\theta(a^t|s^t)}{\pi_{\theta_{'}}(a^t|s^t)} A^t]. \label{eq:sample}
\end{align}

(2) \textit{Policy constraint.} After switching the data sampling policy from $\pi_{\theta}$ to $\pi_{\theta_{'}}$, an issue still remains. Although in the equation (\ref{eq:sample}), they have the similar expectation value of their objective functions, their variances are starkly distinct. Therefore, a KL-divergence penalty can be added as a constraint to the reward formulation to constrain the distances. However, the KL divergence is impractical to calculate in practice as this constraint is imposed on every observation. Thus, we rewrite the objective function as:
$\mathbb{E}_{(s^t,a^t)\sim\pi_{\theta_{'}}}[f^t(\theta)A^t]$~\cite{PPO},
where
\begin{align}
    f^t(\theta)=min\{r^t(\theta), clip(r^t(\theta), 1-\epsilon, 1+\epsilon)\}.
\end{align}

And $r^t(\theta)=\frac{\pi_\theta(a^t|s^t)}{\pi_{{\theta'}}(a^t|s^t)}$. The problem is solved by gradient ascent, therefore, the gradient can be written as:
\begin{align}
    \Delta\theta = \mathbb{E}_{(s^t,a^t)\sim\pi_{\theta_{'}}}[\triangledown f^t(\theta)A^t]. \label{eq:actorobj}
\end{align}

In terms of the value network (Critic), PPO uses identical Critic as per other Actor-Critic algorithms; and the loss function can be formulated in~\cite{PPO} as:
\begin{align}
    L(\phi) = [V_\phi(s^t)-(A^t+V_{\phi'}(s^{t}))]^2. \label{eq:criticloss}
\end{align}

$V(s)$ is the widely used state-value function~\cite{RLintro}, which is estimated by a learned critic network with parameter $\phi$. We update $\phi$ by minimizing the $L(\phi)$, and the parameter $\phi'$ of the target state-value function periodically with $\phi$. Using target value is a prevailing trick in RL, which has been used in many algorithms~\cite{RLintro}.

\subsubsection{Hybrid Reward Architecture (HRA)}

High-dimensional objective functions are common in communication problems, especially for multi-user scenarios, since we usually need to consider multiple factors and distinct requirements of different users. This issue of using RL to solve a high dimensional objective function was first studied in~\cite{HRA}. In their work, they proposed the HRA structure for Deep Q-learning (DQN) which aims to decompose high-dimensional objective functions into several simpler objective functions. HRA has remarkable performance in handling high-dimensional objectives, which serves as the inspiration for our work.

\begin{figure}[t]
\centering
\includegraphics[width=0.75\linewidth]{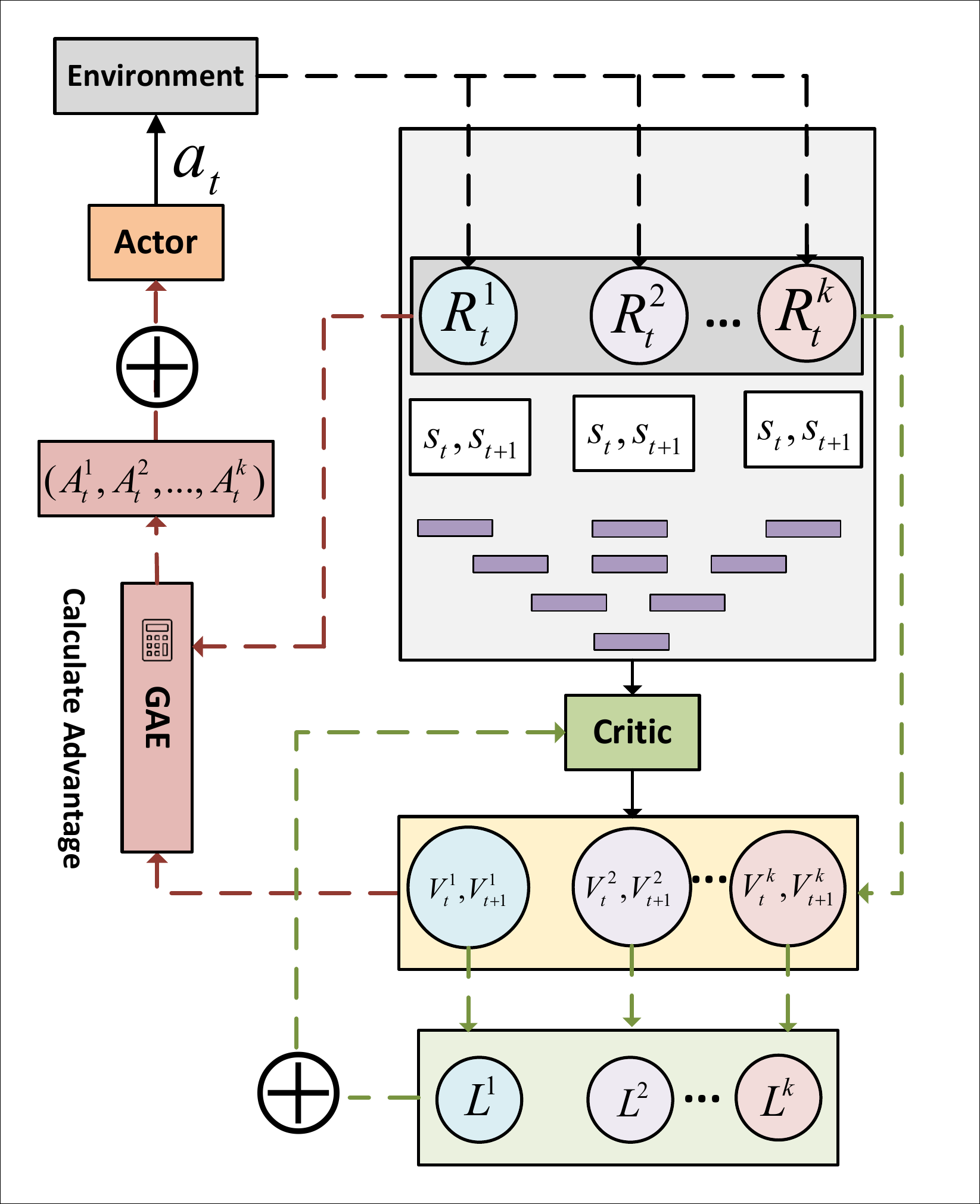}
\caption{Hybrid Reward PPO.}\vspace{-0.7cm}
\label{fig:alg}
\end{figure}

\subsection{HRPPO}
In contrast to decomposing the overall reward into separate sub-goal rewards as done in~\cite{HRA}, we built a user-centered reward decomposition architecture as an extension to PPO, Hybrid Reward PPO (HRPPO), which takes in the rewards of different users and calculates the actions-values separately. In other words, we give the network a view of the state-value of each user, instead of merely evaluating the overall value of an action based on an overall state-value.

\textbf{Function process:} In each episode, when the current transmission is accomplished with the selected action $a_t$, the environment will issue rewards $R_1^t, R_2^t,..., R_n^t$ as feedback to different VUs. These rewards along with their corresponding states and next iteration state, will be sent to the Critic to generate the state-values $V_1^t, V_2^t, ..., V_n^t$, representing the state-value of each VU. The state-value is then used to calculate the advantages and losses for each VU. The above-mentioned process is illustrated in Fig. \ref{fig:alg}.

\textbf{Update function:} In equation~(\ref{eq:actorobj}), we established the policy gradient for PPO Actor, and in HRPPO we have the gradient $\Delta\theta$ as:
\begin{align}
    &\Delta\theta = \mathbb{E}_{(s^t,a^t)\sim\pi_{{\theta}'}}[\triangledown f^t(\theta,(\sum_{n=1}^N A_n^t(s^t))].
    \label{eq:gradient}
\end{align}
where $A_n^t$ denotes the advantages of different VUs. The generalized advantage estimation (GAE)~\cite{schulman2015high} is chosen as the advantage function:
\begin{align}
    &A_n^t = \delta^t + (\gamma\lambda)\delta_n^{t+1}+...+(\gamma\lambda)^{\bar{T}-1}\delta_n^{t+\bar{T}-1}, \\
    &\text{where}~~~\delta_n^t=R_n^t+\gamma V_{\phi'}(s^{t+1})-V_{\phi'}(s^t).
\end{align}
$\bar{T}$ specifies the length of the given trajectory segment, $\gamma$ specifies the discount factor, and $\lambda$ denotes the GAE parameter.
In terms of Critic loss, the equation~(\ref{eq:criticloss}) is formatted into:
\begin{align}
    L(\phi) = \sum_{n=1}^N \left (V_{\phi,n}(s^t)-(A_n^t + V_{\phi',n}(s^{t})) \right )^2. \label{eq:HRloss}
\end{align}
Similar to the renowned centralized training decentralized execution (CTDE) framework~\cite{CTDE}, the $V_\phi^n$ also uses centralized training with equation~(\ref{eq:HRloss}). Therefore, the training time will not scale with the number of users.

\subsubsection{\textbf{Baselines}}
We also implement some of the most renowned RL algorithms that are capable of tackling problems with a discrete action space.

\begin{itemize}
    \item \textbf{HRDQN}. We implemented the hybrid reward DQN following the structure of HRA~\cite{HRA}.
    \item \textbf{PPO}. The traditional PPO is used as a baseline. The sum of all users' rewards is selected as the global reward.
    \item \textbf{Random}. The random Agent selects actions randomly, which represents the system performance if no channel resource allocation is performed.
\end{itemize}
\vspace{-0.2cm}

\subsection{Metrics}
We introduce a set of metrics (apart from RL rewards) to evaluate the effectiveness of our proposed methods.

\begin{itemize}
    \item \textbf{Successful frames}. The number of successful frames among total $T$ frames determines the Frame Rate per Second (FPS) of the virtual world scenes, and hence fluidity of the Metaverse VR experience.
    \item \textbf{Energy consumption}. We illustrate the total energy consumption in each episode. Lower energy consumption signifies a more effective use of channel resources.
    \item \textbf{Average rate}. The average downlink transmission rate of all VUs and frames in each episode is shown to evaluate the trained policy. A higher average rate indicates better allocation of channel resources.
\end{itemize}
\vspace{-0.2cm}

\begin{figure}[!t] 
        \renewcommand{\algorithmicrequire}{\textbf{Initiate:}}
        \renewcommand{\algorithmicensure}{\textbf{Output:}}
        \begin{algorithm}[H]
            \caption{\label{alg:PPO} HRPPO}
            \begin{algorithmic}[1]
                \REQUIRE Actor $\theta$, Critic $\phi$ and target network $\phi'$
                \FOR{iteration = $1,2...$}
                    \STATE $Agent$ execute action according to $\pi_{\theta^{'}}(a^t|s^t)$
                    \STATE Get reward $R_1^t,R_2^t,...,R_N^t$ and next state $s^{t+1}$
                    \STATE $s^t \leftarrow s^{t+1}$ 
                    \STATE Sample ($s^t, a^t, (R_1^t,R_2^t,...,R_n^t), s^{t+1}$) till end
                    \STATE Compute advantages \{$A_1^t,...,A_N^t$\} and target values \{$V_{1,targ}^t,...,V_{N,targ}^t$\} using current Hybrid Critic 
                    \FOR{$k$ = $1,2,...,K$}
                        \STATE Shuffle the data's order, set batch size $bs$
                        \FOR{$j$=$0,1,...,\frac{T}{bs}-1$}
                            \STATE Compute gradient for Actors by eq.~(\ref{eq:gradient})
                            \STATE Update Actors by gradient ascent
                            \STATE Update Critic with MSE loss using eq.~(\ref{eq:HRloss})
                        \ENDFOR
                        \STATE Assign target network $\phi' \leftarrow \phi$ every $C$ steps
                    \ENDFOR
                \ENDFOR
            \end{algorithmic}
        \end{algorithm}\vspace{-0.8cm}
\end{figure}

\section{Experiment results}

\subsection{Numerical Setting}
Consider a $30\times30$ $m^2$ indoor space where multiple VUs are distributed uniformly across the space. We set the number of channels to be $3$ in each experiment configuration, and the number of VUs to be from $5$ to $8$ across the different experiment configurations. The maximum resolution of one frame is 2k ($2048\times1080$) and the minimum is 1080p ($1920\times1080$). Each pixel is stored in 16 bits~\cite{oculus} and the factor of compression is 150~\cite{VRsurvey}. We randomize the data size of one frame to take values from a uniform distribution in which $D_n^t\in[\frac{1920\times1080\times16}{150}, \frac{2048\times1080\times16}{150}]$. The flashed rate, $T$ frames in one second is taken to be 90, which is considered the best rate for VR~\cite{VRsurvey} applications. The bandwidth of each channel is set to $10\times180$ kHZ. The required successful frame transmission count $\tau_{n,F}$ is uniformly selected from $[75,80]$, which is higher than the acceptable of $60$~\cite{VRsurvey}. In terms of channel gain, the small-scale fading follows the Rayleigh distribution and $\alpha=2$ is the path loss exponent. For all experiments, we use $2\times 10^5$ steps for training, and the evaluation interval is set to be $50$ training steps. As there are several random variables in our environment, all experiments are conducted under global \textbf{random seeds from 0-10}, and the error bands are drawn to better illustrate the model performances.

\label{experiment}

\begin{figure*}[t]
\centering
\subfigtopskip=2pt
\subfigbottomskip=2pt

\subfigure[Training reward with 6 VUs.]{
\begin{minipage}[t]{0.24\linewidth}
\centering
\includegraphics[width=1\linewidth]{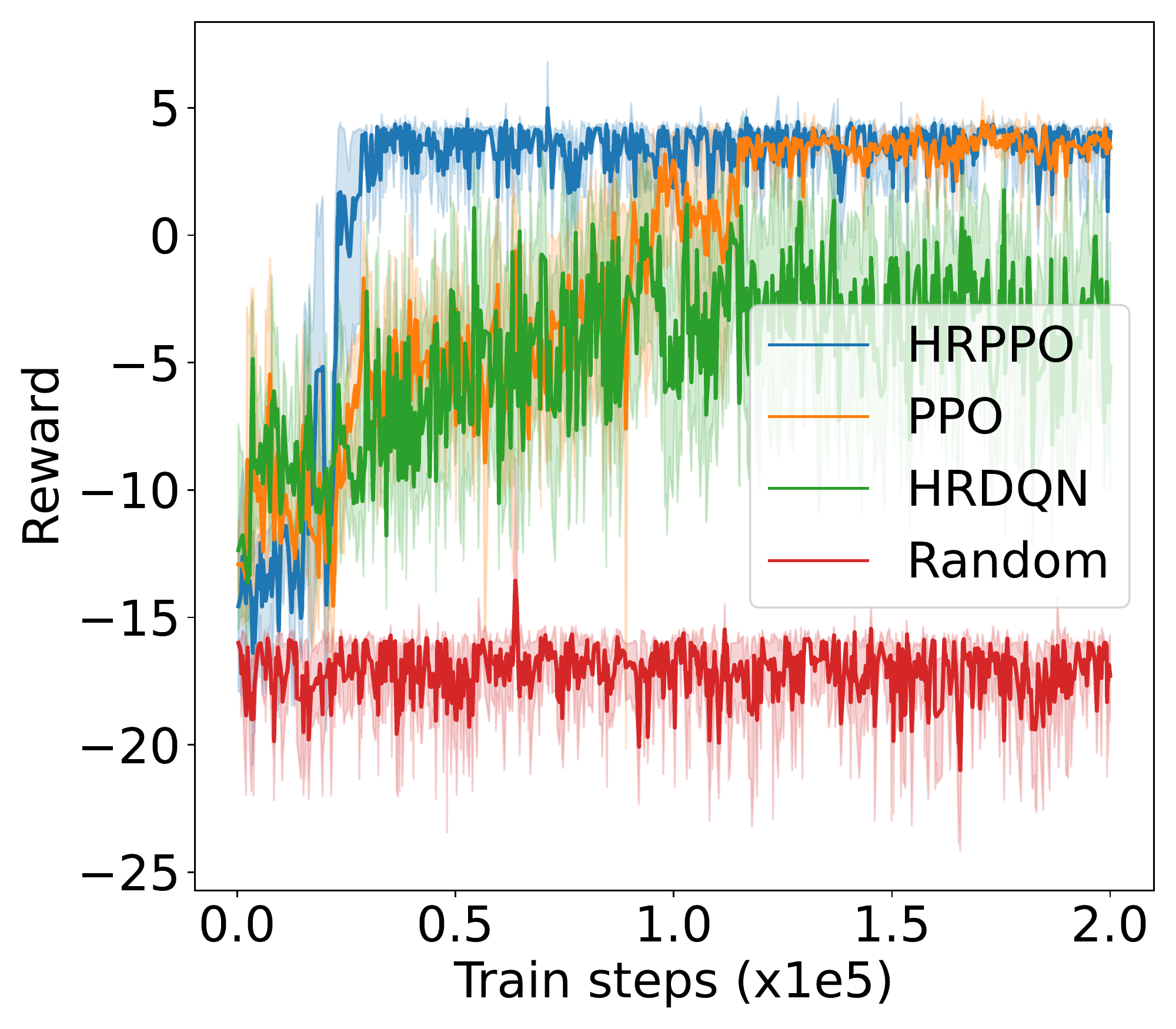}
\label{fig:83}
\vspace{-10mm}
\end{minipage}
}%
\subfigure[Successful frames with 6 VUs.]{
\begin{minipage}[t]{0.24\linewidth}
\centering
\includegraphics[width=1\linewidth]{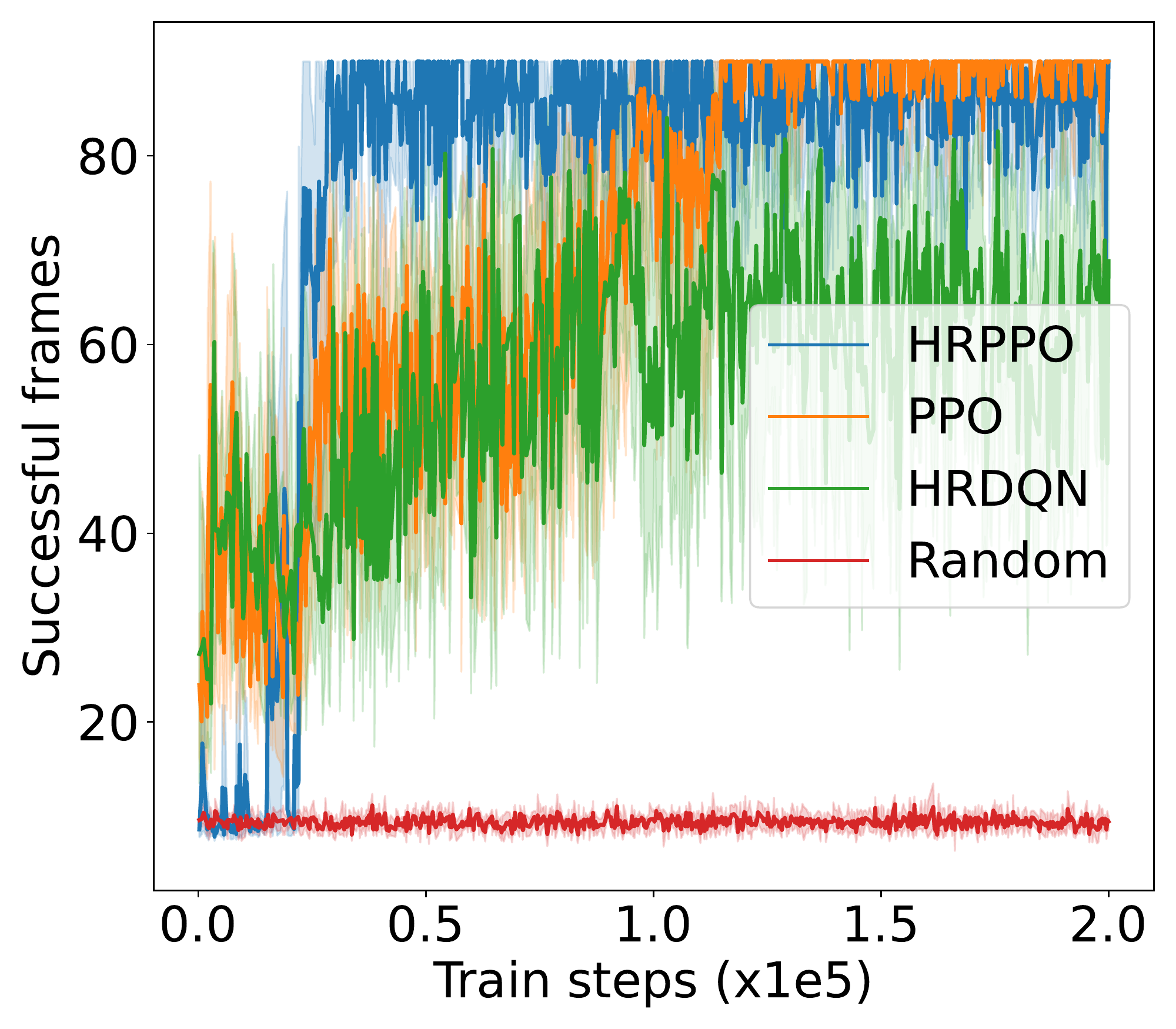}
\label{fig:ave}
\vspace{-10mm}
\end{minipage}
}%
\subfigure[Energy consumption with 6 VUs.]{
\begin{minipage}[t]{0.24\linewidth}
\centering
\includegraphics[width=1\linewidth]{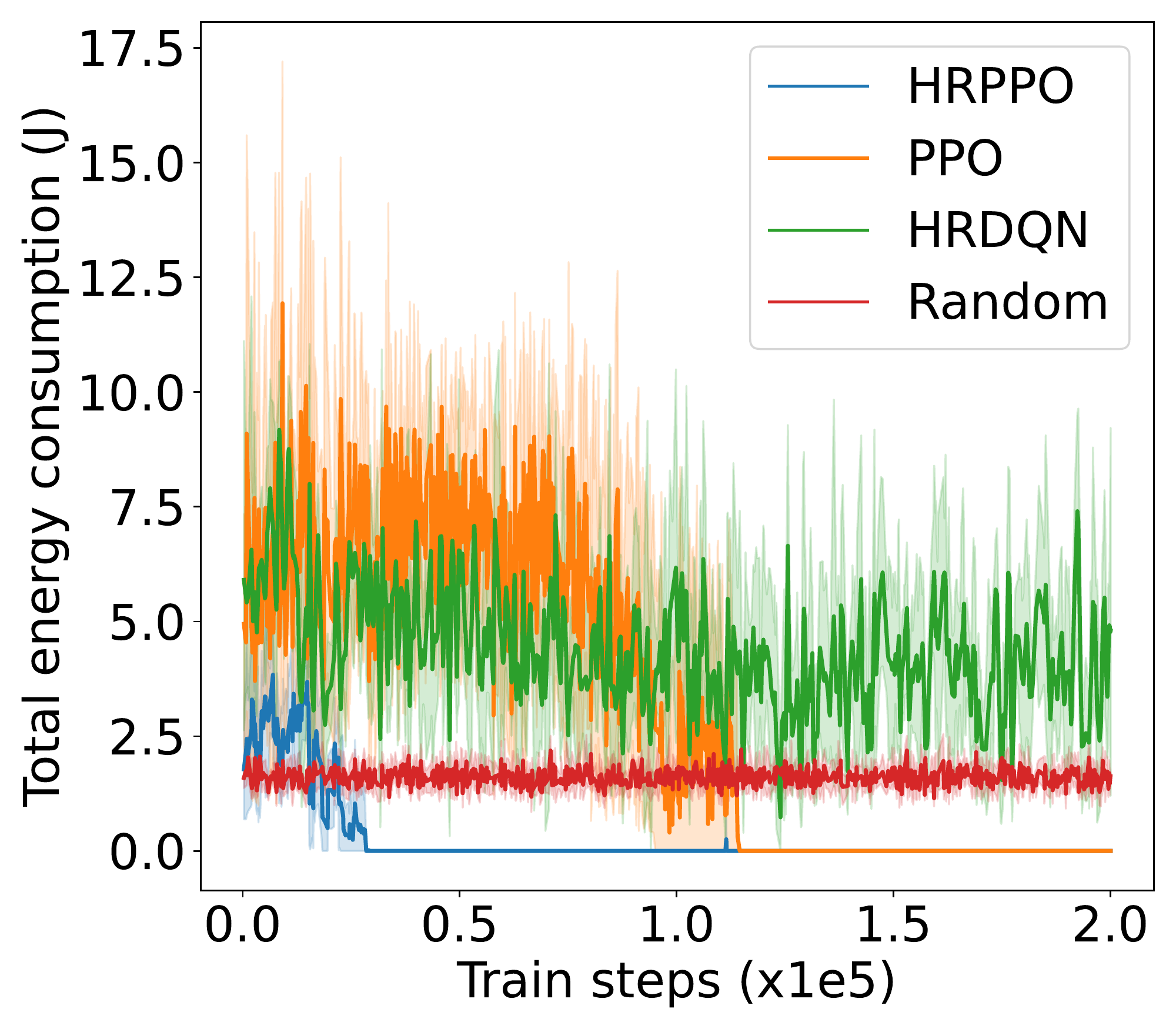}
\label{fig:43}
\vspace{-10mm}
\end{minipage}%
}%
\subfigure[Average rate with 6 VUs.]{
\begin{minipage}[t]{0.24\linewidth}
\centering
\includegraphics[width=1\linewidth]{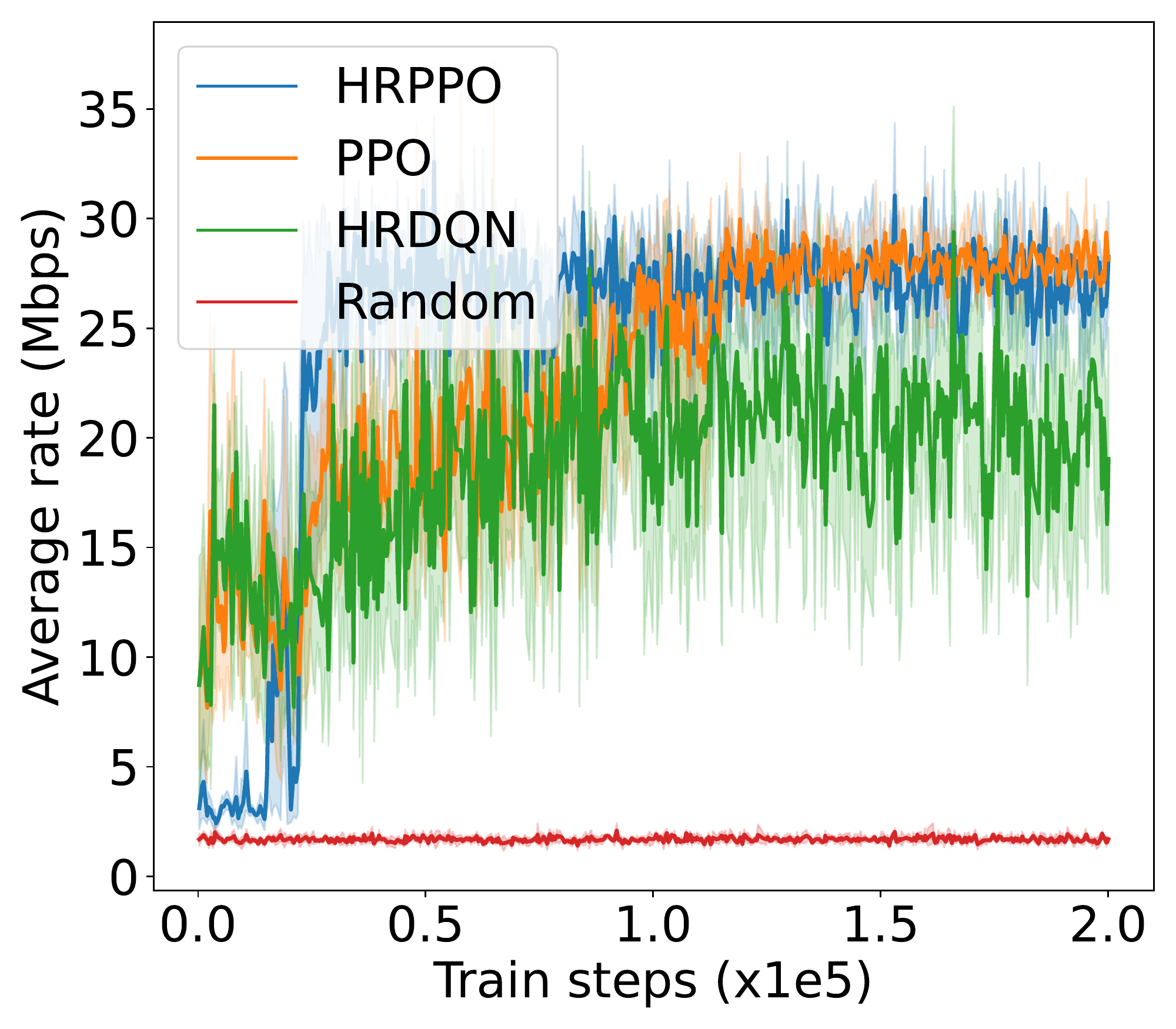}
\label{fig:63}
\vspace{-10mm}
\end{minipage}%
}%

\subfigure[Training reward with 8 VUs.]{
\begin{minipage}[t]{0.24\linewidth}
\centering
\includegraphics[width=1\linewidth]{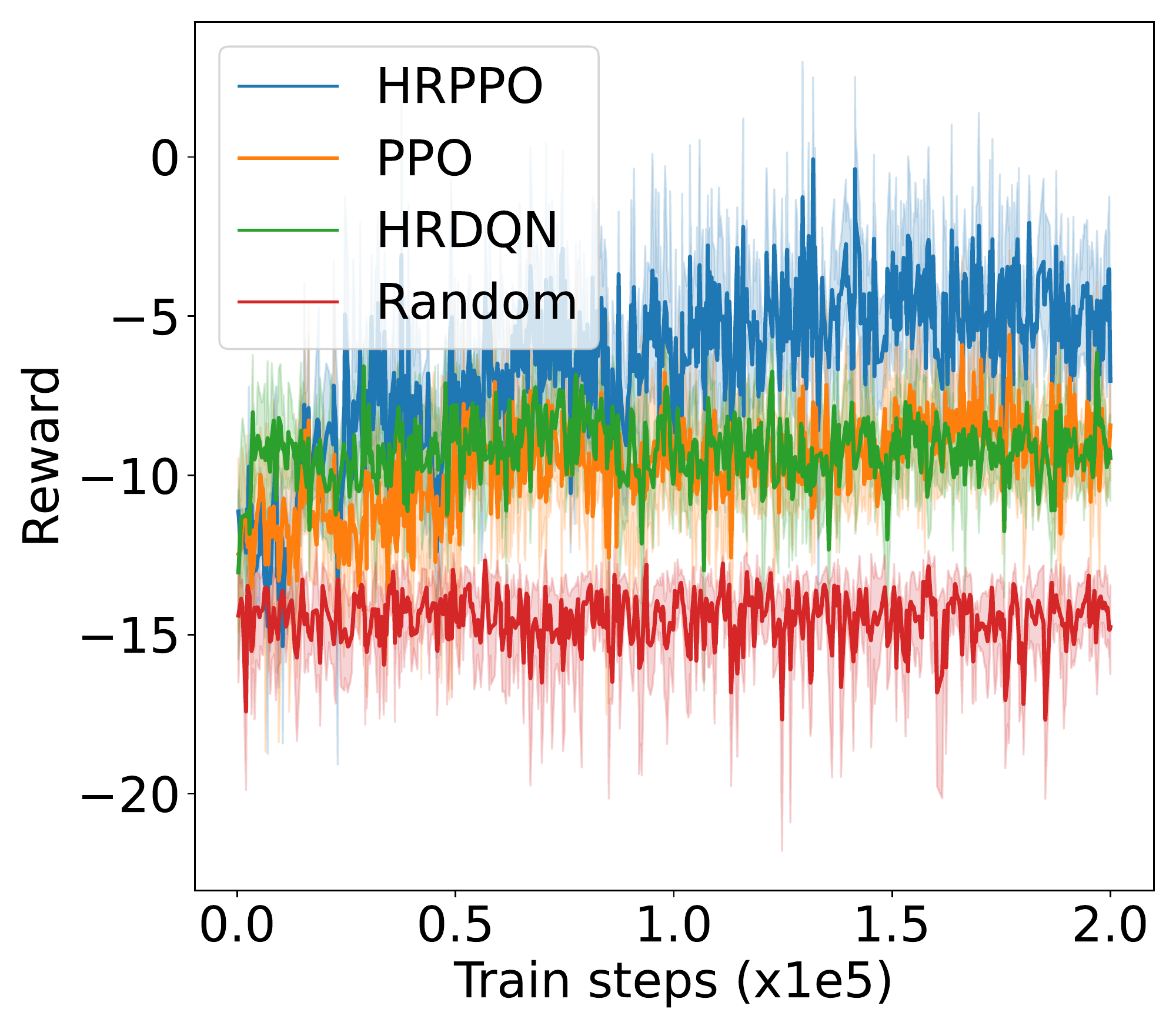}
\label{fig:38reward}
\vspace{-10mm}
\end{minipage}
}%
\subfigure[Successful frames with 8 VUs.]{
\begin{minipage}[t]{0.24\linewidth}
\centering
\includegraphics[width=1\linewidth]{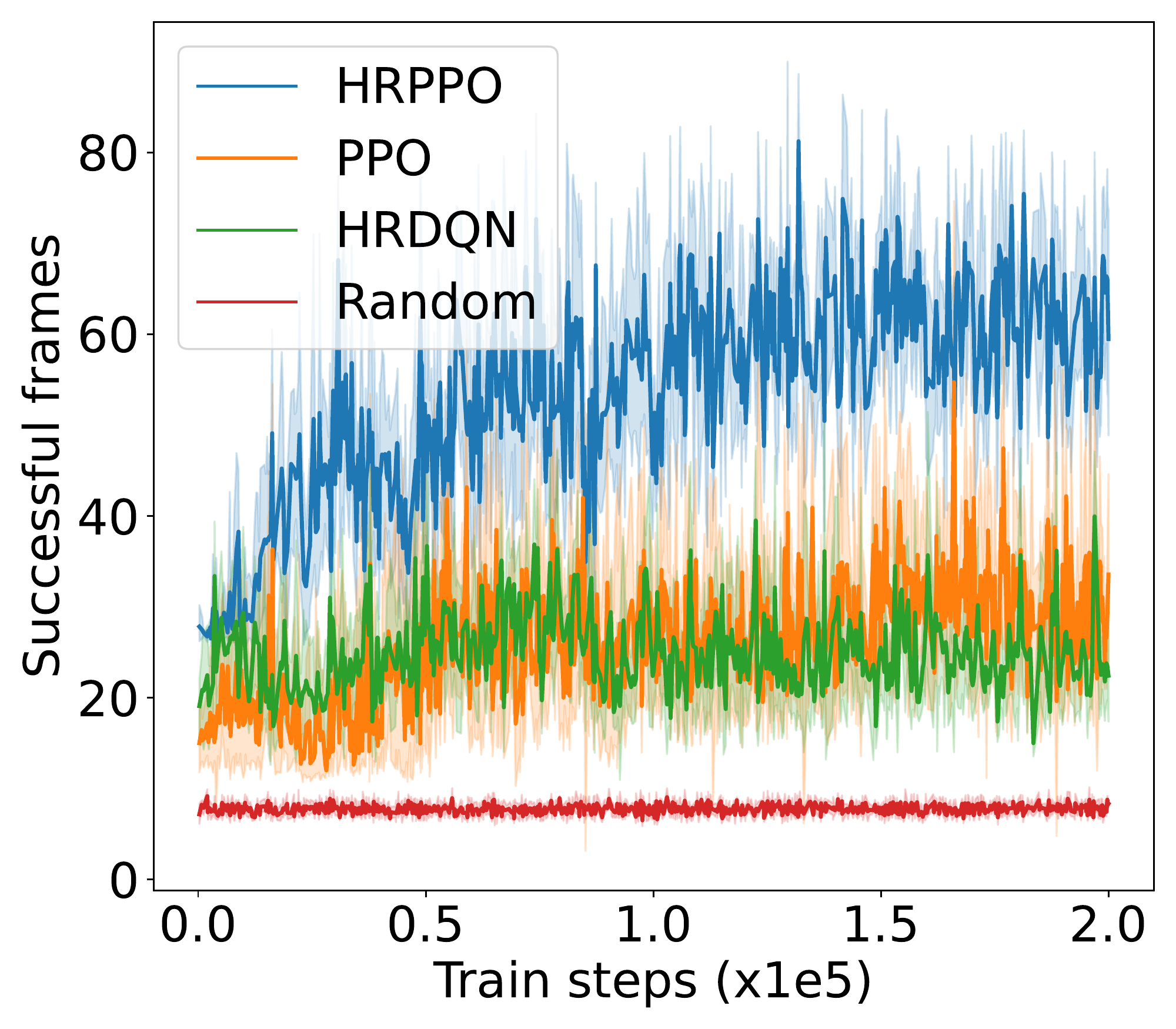}
\label{fig:38step}
\vspace{-10mm}
\end{minipage}
}%
\subfigure[Energy consumption with 8 VUs.]{
\begin{minipage}[t]{0.24\linewidth}
\centering
\includegraphics[width=1\linewidth]{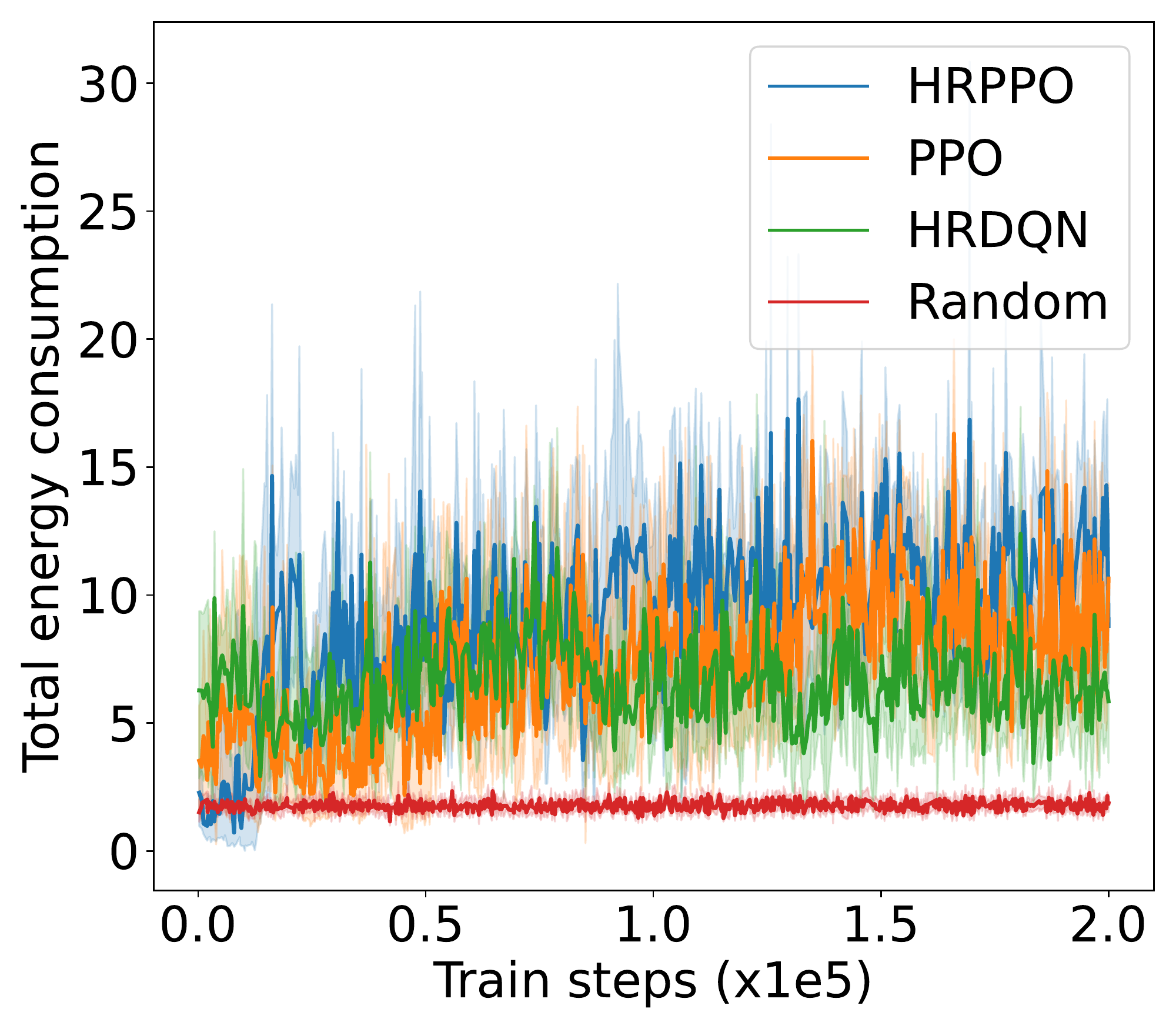}
\label{fig:38energy}
\vspace{-10mm}
\end{minipage}%
}%
\subfigure[Average rate with 8 VUs.]{
\begin{minipage}[t]{0.24\linewidth}
\centering
\includegraphics[width=1\linewidth]{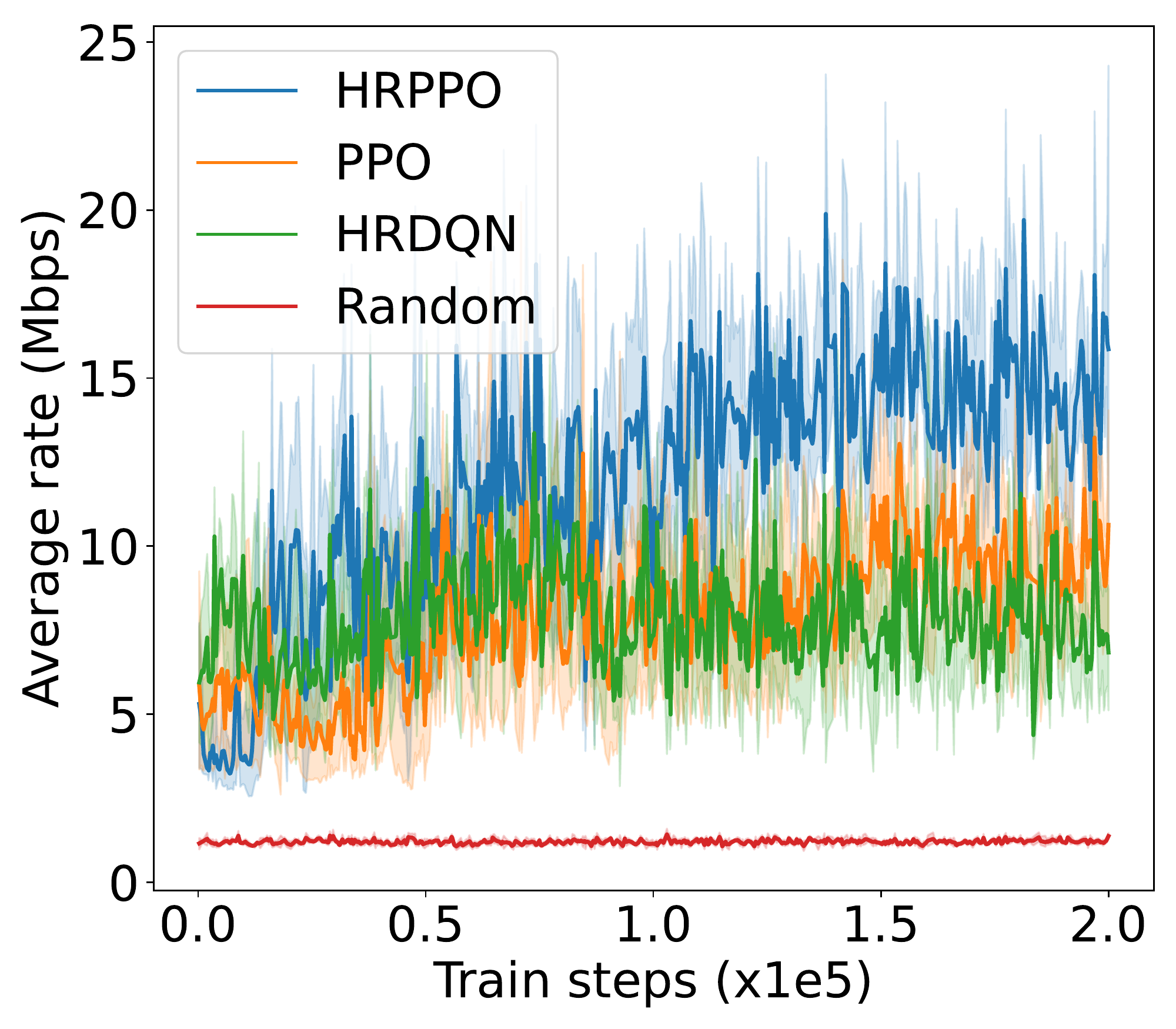}
\label{fig:38rate}
\vspace{-10mm}
\end{minipage}%
}%

\caption{Metrics in environments with 6 VUs and 8 VUs. Considering the randomly evolving environment, all experiments are conducted with global random seeds from 0-10, and the error bands are drawn.}
\label{fig:train}
\vspace{-0.5cm}
\end{figure*}

\subsection{Result analysis}
We first illustrate the performances of the different models against different metrics in two experimental configurations (shown in Fig.~\ref{fig:train}): one with 6 VUs and the other with 8 VUs. We then show the overall results for each experimental configurations in Table ~\ref{table:results}. Results in Table ~\ref{table:results} are taken from the average of the final $200$ steps.

The training reward, successful frame transmission counts, and average downlink transmission rate show an overall upward trend as training progresses. When pitted against these metrics, HRPPO performed the best out of the tested baseline algorithms. In the experimental setting with $6$ VUs, although PPO and HRPPO are able to attain similar peak rewards towards later training stages, HRPPO converges in half the number of training steps taken for PPO to achieve convergence. In the experimental setting with $8$ VUs, HRPPO obtains a much higher final reward when compared to PPO. Both HRPPO and PPO achieved higher rewards than HRDQN, and performed better in each metric. HRPPO and PPOs' performance superiority can be attributed to the PPO's policy KL penalty and higher sample efficiency. However, HRDQN and PPO fail to find a good solution in more complicated scenarios. In the $6$ VU experimental setting, both HRPPO and PPO are able to allocate VUs to a VSP channel for computation offloading in each round, and this is reflected in zero energy spent on local device computation. However, in the experimental setting with $8$ VUs, there is insufficient channel resources and all of the three algorithms learn strategies to increase transmission rate and avoid frame rate decrement, by allocating some VUs to perform local computation, which increases energy consumption. 


The complete results in Table.~\ref{table:results} shows that HRPPO obtains the best performance for almost every metric under every scenario. This demonstrates that decomposing the reward and using sum-losses which provides a user-centered view to the RL agent, is good approach to tackling a multi-user computation offloading problem.

\begin{table}[t]
\centering
\caption{Overall results}
\label{table:parameter}
\vspace{-0.2mm}
\scalebox{0.9}{
\begin{tabular}{ccccc}
\hline
VU number & Reward & \makecell{Successful\\frames} & \makecell{Energy\\consumption(J)} & \makecell{Average\\rate (Mbps)} \\ \hline
\multicolumn{5}{c}{HRPPO} \\ \hline
$5$ & \textbf{4.66} & \textbf{88.44} & \textbf{0} & \textbf{28.33}\\
$6$ & \textbf{3.26} & \textbf{88.35} & \textbf{0} & \textbf{27.56}\\
$7$ & \textbf{-2.21} & \textbf{83.67} & \textbf{7.05} & \textbf{19.90} \\
$8$ & \textbf{-5.15} & \textbf{65.69} & $11.57$ & \textbf{15.02}\\ \hline
\multicolumn{5}{c}{PPO} \\ \hline
$5$ & $4.41$ & $88.02$ & \textbf{0} & $27.43$\\
$6$ & $3.18$ & $86.27$ & \textbf{0} & $26.62$ \\
$7$ & $-5.97$ & $61.42$ & $7.08$ & $15.01$ \\
$8$ & $-9.02$ & $29.38$ & \textbf{9.12} & $9.92$\\ \hline
\multicolumn{5}{c}{HRDQN} \\ \hline
$5$ & $1.13$ & $79.38$ & $1.05$ & $23.84$\\
$6$ & $-4.27$ & $63.92$ & $4.45$ & $19.91$\\
$7$ & $-7.03$ & $44.19$ & $7.88$ & $13.61$\\
$8$ & $-9.22$ & $25.85$ & $7.03$ & $8.03$\\ \hline
\end{tabular}
}
\label{table:results}
\vspace{-0.5cm}
\end{table}

\section{conclusion}
\label{conclude}
In this paper, we study a multi-user VR in the Metaverse mobile edge computing over wireless networks scenario. Multiple users with varying requirements are considered, and a novel user-centered RL algorithm \textit{HRPPO} is designed to tackle it. Extensive experiment results show that \textit{HRPPO} has the quickest convergence and achieves the highest reward, which is $45\%$ higher than the traditional \textit{PPO}. In the future, we will continue to optimize the power allocation to seek more optimal solutions to our proposed problems.

\section*{Acknowledgement}

This research is partly supported by Singapore Ministry of Education Academic Research Fund under Grant Tier 1 RG90/22, RG97/20, Grant Tier 1 RG24/20 and Grant Tier 2 MOE2019-T2-1-176; partly by the NTU-Wallenberg AI, Autonomous Systems and Software Program (WASP) Project.

{\small
\bibliographystyle{IEEEtran}

}

\end{document}